\documentclass[12pt]{article}

\usepackage{graphics,a4wide,amssymb}

\newcommand{\be}{\begin{equation}}
\newcommand{\ee}{\end{equation}}
\newcommand{\bea}{\begin{eqnarray}}
\newcommand{\eea}{\end{eqnarray}}
\newcommand{\bref}[1]{(\ref{#1})}
\newcommand{\nsp}{\hspace{-0.2in}}

\newcommand{\vp}{\varphi}
\newcommand{\dGS}{\delta_{\scriptscriptstyle GS}}
\newcommand{\sR}{s_{\!\scriptscriptstyle R}}
\newcommand{\g}{\gamma}
\newcommand{\mP}{m_\mathrm{Pl}}
\newcommand{\GeV}{\, \mathrm{GeV}}
\newcommand{\TeV}{\, \mathrm{TeV}}
\newcommand{\eV}{\, \mathrm{eV}}

\begin{document}

\title{\bf Local Axion Cosmic Strings from Superstrings}
\author{
Stephen C. Davis$^a$\footnote{Stephen.Davis@epfl.ch},
Pierre Bin\'etruy$^{b,c}$\footnote{Pierre.Binetruy@th.u-psud.fr} 
\ and Anne-Christine Davis$^d$\footnote{A.C.Davis@damtp.cam.ac.uk} \\ \\
\em ${}^a$Institute of Theoretical Physics, \'Ecole Polytechnique
F\'ed\'erale de \\ \em  Lausanne, CH--1015 Lausanne, Switzerland \\
\em ${}^b$LPT, Universit\'e de Paris-Sud, B\^at. 210,
91405 Orsay Cedex, France \\
\em ${}^c$APC, Universit\'e Paris VII, 2 place Jussieu, 
75251 Cedex 05, France \\
\em ${}^d$Department of Applied Mathematics and Theoretical Physics, 
Centre for \\ \em Mathematical Sciences, University of Cambridge,
Cambridge, CB3 0WA, UK
}

\maketitle

\begin{abstract}
Axionic cosmic string solutions are investigated in a superstring
motivated model with a pseudo-anomalous $U(1)$ gauge symmetry. The
inclusion of a gauge field and spatially varying dilaton allow local defect
solutions with finite energy per unit length to be found. Fermion zero
modes (whose presence is implied by supersymmetry) are also
analysed. The corresponding fermion currents suggest strong cosmological
bounds on the model. It is shown that the unusual form of
the axion strings weakens these bounds. Other cosmological constraints
on the underlying theory are also discussed.
\end{abstract}

\section{Introduction}

Spontaneous breaking of a pseudo-anomalous $U(1)$ gauge symmetry
occurs in a large class of superstring
compactifications~\cite{supstr}. The $U(1)$ symmetry arises
generically as a remnant of the Green-Schwarz~\cite{GS} mechanism of
anomaly cancellation in the underlying ten-dimensional supergravity.

It is well known that line-like topological defects form during an
ordinary $U(1)$ symmetry breaking. Such defects are called cosmic
strings, and have many interesting cosmological
properties~\cite{CS}. Similar defects can also form if the symmetry is
pseudo-anomalous. These strings are expected to be global, because of
their coupling to the axion~\cite{oldaxion}. Furthermore, it seems
that a cutoff must be introduced inside the string core in order to
give a finite energy per unit length. It has been shown that by
coupling the axion to a gauge field, local axion strings can be
obtained (although the cutoff inside the core is still
needed)~\cite{ax1}. The coupling of the dilaton-axion field to the
gauge fields plays a central role in the Green-Schwarz mechanism, so
local axion strings are natural in this context.

Recent work on the formation of topological defects in superstring
theory has led to a renewed interest in cosmic
strings~\cite{Dstrings}. In contrast to other work, we will just be
considering a four-dimensional model, although we expect our results
to be of general relevance.

In this paper we discuss axion string solutions which do not require a
cutoff in the core. In a supersymmetric theory, the dilaton
forms part of the same supermultiplet as the axion. By allowing the
dilaton to vary inside the string, finite energy solutions can be
found.  The axion field varies (in a topologically non-trivial way)
around the string, and so plays the role of the phase of the Higgs
field in a conventional $U(1)$ cosmic string model. In a similar way
the dilaton plays the role of the magnitude of the Higgs field.  This
type of solution relies on particular couplings between the axion and
dilaton fields, although they are completely natural in a
supersymmetric (or superstring) context.

Although we have referred to the dilaton in the above discussion, the same
ideas apply for pseudo-anomalous $U(1)$ breaking with an axion-moduli
superfield. Similarly the analysis and results of this paper will also
apply to moduli, although the corresponding couplings and mass scales
coming from the underlying theory will be different.

We will investigate local, axion strings in a superstring motivated,
supersymmetric model. A simplified version of a `racetrack'
potential~\cite{krasnikov} will be used to stabilise the dilaton. Such
a potential could arise from gaugino condensation. It is quite natural
for its energy scale to be far lower than that associated with
supergravity, perhaps even as low as the QCD scale.  For simplicity we
model these lower energy effects with two effective scalar fields. The
model is outlined in section~\ref{sec:model}, and the cosmic string
solutions themselves are discussed in section~\ref{sec:string}.

A common feature of supersymmetric cosmic string models is the
existence of conserved, massless fermion currents which flow along the
strings~\cite{sscs}. Like the strings themselves, these have
interesting cosmological implications~\cite{witten}. The massless
fermion bound states of our solutions are examined in
section~\ref{sec:fermion}. The cosmological evolution of a network of
such strings is discussed in section~\ref{sec:network}. The bounds on
the underlying theory coming from the cosmological properties of the
axion strings are derived in section~\ref{sec:cosmo}. The possibility
of cosmic string loops which are stabilised by the fermion currents
generally gives rise to particularly strong bounds. However, as we
will show, these bounds are significantly weakened for our string
solutions. 

In section~\ref{sec:bog} we discuss an apparent analytic
simplification of the field equations, and in section~\ref{sec:con} we
summarise our results.

\section{Axion cosmic string model}
\label{sec:model}

We will consider a supersymmetric model with a pseudo-anomalous $U(1)$
gauge symmetry. Working in units with $\mP =1$, the
Lagrangian is
\be
\tilde \mathcal{L} = 
\left(\Phi_i^\dagger e^{2q_i V} \Phi_i + \mathcal{K} \right)
+ \left( \frac{1}{4} S W^\alpha W_\alpha 
+ W(\Phi_i,S) \right) \delta(\bar\theta^2) + (\mbox{h.\ c.}) \ .
\label{LagS}
\ee
$S(s, 2\sR \chi_\alpha, F_s)$ is the axion-dilaton superfield,
$V(A_\mu, \sR^{-1/2} \lambda_\alpha, D)$ is the gauge vector superfield, and
$\Phi_i(\phi_i, \psi_{i\alpha}, F_i)$ are chiral superfields with
charges $q_i$. The kinetic term for the superfield $S$ is described by the
modified K\"ahler function
$\mathcal{K}= -\log (S + \bar S - 4\dGS V)$, where $\dGS$ is the Green-Schwarz
parameter~\cite{supstr}.

If the dilaton is to have a finite vacuum expectation value, a
non-trivial superpotential ($W$) is required. We will consider a
racetrack~\cite{krasnikov} style model with the effective superpotential
\be
W = \Phi_0 \left\{h_1 (\Phi/\eta)^{\! n_1} e^{-3S/(2b_1)}
- h_2 (\Phi/\eta)^{\! n_2} e^{-3S/(2b_2)} \right\}
\ee
where $b_i = 3N_i/(16\pi^2)$. The integers $N_i$ are
determined by the symmetry breaking of the original theory. The above
superpotential resembles that arising from an $E_8 \to SU(N_1) \times
SU(N_2)$ breaking, in which case $N_1+N_2 \leq 10$.

The effective superfields, $\Phi$ and $\Phi_0$, represent
superstring and gaugino condensation effects. We take $q_\phi = -1$
and $q_0<0$. Gauge invariance implies that $q_0 = n_i - 3\dGS/b_i$,
and so for our model we require
\be
\dGS  = \frac{N_1N_2(n_1 -n_2)}{16 \pi^2(N_2-N_1)} \ . 
\ee
Typically we expect $\dGS \sim 1/10$~\cite{supstr},
but we will consider more general values.

The Lagrangian \bref{LagS} can be expanded in terms of the component
fields. The auxiliary fields can then be eliminated using their
equations of motion. Working in Wess-Zumino gauge, the bosonic part of the
Lagrangian is
\bea
\mathcal{L}_\mathrm{B} &=&  \frac{1}{4\sR^2} \partial_\mu \sR \partial^\mu \sR
+ \frac{1}{4\sR^2} \left(\partial_\mu a - 2\dGS A_\mu \right)^2
- \frac{\sR}{4} F^{\mu \nu} F_{\mu \nu}  
+ \frac{a}{4} F^{\mu \nu} \tilde F_{\mu \nu}
\nonumber \\ && {}
+ |\mathcal{D}_{\! \mu} \phi|^2 
+ |\mathcal{D}_{\! \mu} \phi_0|^2
- |F_0|^2 - |F|^2 - {1 \over 4\sR^2} |F_s|^2 - {\sR \over 2} D^2
\label{LagWZ}
\eea 
where $s = \sR + ia$, $\mathcal{D}_{\! \mu}\phi = (\partial_\mu-i A_\mu)\phi$, 
and $\mathcal{D}_{\! \mu}\phi_0 = (\partial_\mu + iq_0 A_\mu)\phi_0$. 
The real part of the dilaton defines the gauge coupling $1/g^2 = \sR$. 
The auxiliary fields are given by
\be
F_0^\dagger = -h_1 (\phi/\eta)^{n_1} e^{-3s/(2b_1)} 
+  h_2 (\phi/\eta)^{n_2} e^{-3s/(2b_2)}
\ee
\be 
D = -{1 \over \sR}\left(q_0 |\phi_0|^2 - |\phi|^2 + {\dGS \over \sR}\right)
\ .
\ee
$F$ and $F_s$ are both proportional to $\phi_0$. The scalar potential 
is minimised by $\sR=\infty$ or
\be
\phi_0=0 \ , \ \ |\phi| = \eta \ , \ \ 
\sR = \frac{1}{g_0^2} = \frac{\dGS}{\eta^2}\ ,
\ee
where $g_0$ is the vacuum value of the coupling $g$.
The superpotential is related to $g_0$ by
\be
\frac{2\dGS}{n_1-n_2} \ln \frac{h_1}{h_2} = \frac{1}{g_0^2} \ .
\ee

We see that the model has two distinct mass scales. One, 
\be
m_D = \frac{\sqrt{2} \eta^2}{\sqrt{\dGS}} \sqrt{1+\eta^2}
= \sqrt{2\dGS} g_0^2 \sqrt{1+\eta^2} \ ,
\ee
arises from the $D$ term, and because of supersymmetry is also the mass
of the gauge field $A_\mu$. 
If $\dGS \sim 1/10$ and $g_0 \sim 1$, as suggested by string theory, we
will have $m_D \approx 1/\sqrt{5}$. The other mass scale,
\bea
m_F =&&\nsp |n_2-n_1| \frac{|h_1|}{\eta^2}
\exp\left(-\frac{3\dGS}{2b_1 \eta^2}\right) \sqrt{1+\eta^2} 
\nonumber \\
=&&\nsp 
16 \pi^2 \frac{|h_1|}{g_0^2} \frac{|N_1-N_2|}{N_1N_2} 
\exp\left(-\frac{8\pi^2}{N_1 g_0^2}\right) \sqrt{1+\eta^2} \ , \hspace{0.3in}
\eea
arises from the $F_0$ term, and will generally be far smaller. If
$\eta$ is small, the fields $s$ and $\phi$ are approximate mass
eigenstates. Their masses are respectively $m_F$ and $m_D$.

\section{Cosmic Strings}
\label{sec:string}

We will look for cosmic string solutions of the form
\bea
\phi &=& \eta f(r) e^{in\vp} \label{ansst} \\
A_\varphi &=&  n \frac{v(r)}{r}  \\
s &=& \frac{\dGS}{\eta^2 \g(r)^2} + 2 i n \dGS \vp \\
\phi_0 &=& 0 \ . \label{ansend}
\eea
Note that $q_0 n$ must be an integer so that $F_0$ is single valued. 
As $r$ approaches infinity $f$, $\g$ and $v$ all tend to 1. The
boundary conditions at $r =0$ are $\g=f=v=0$. In contrast
to previous work~\cite{oldaxion,ax1}, we allow
spatial variations of the dilaton.

The equations of motion for the various fields are
\bea
&&\partial^\mu \left( \frac{1}{\sR} \partial_\mu \sR\right) 
+ {1 \over \sR^2}(\partial_\mu a - 2 \dGS A_\mu)^2 
+ {\sR \over 2} F^{\mu \nu} F_{\mu \nu}
+ 4 \sR F_0 {\partial F_0^* \over \partial s^*} 
\nonumber \\&& \hspace{2in} {}
- \frac{1}{\sR}\left(|\phi^2| - {\dGS \over \sR}\right)
\left(|\phi^2| - 3{\dGS \over \sR}\right) = 0 \hspace{0.4in}
\label{seq}
\eea
\be
\mathcal{D}^\mu \mathcal{D}_{\! \mu} \phi
+ F_0 {\partial F_0^* \over \partial \phi^*} 
+ {1 \over \sR} \left(|\phi^2| - {\dGS \over \sR}\right)\phi = 0
\label{phieq}
\ee
\be
\partial^\mu (\sR F_{\mu \nu})
+ 2 \mbox{Im}[\phi(\partial_\nu+iA_\nu)\phi^*]
- {\dGS \over \sR^2}[\partial_\nu a -2\dGS A_\nu] = 0 \ . 
\label{Aeq}
\ee

The equations for the string profile function are found by substituting
the string ansatz (\ref{ansst}--\ref{ansend}) into the above field
equations. It is useful to introduce $\rho = r \eta^2 \sqrt{2/\dGS}$.
The equations (\ref{seq}--\ref{Aeq}) for the string solution are then
\bea
\frac{1}{\rho} \left(\frac{\rho \g'}{\g}\right)' 
- 2 n^2\eta^4 \frac{(1-v)^2}{\rho^2} \g^4
+ n^2 \eta^2 \g^2 \left(\frac{v'}{\g^2 \rho}\right)^2
- \frac{\eta^2}{4}(f^2 -\g^2)(f^2 - 3\g^2)\g^2
\hspace{0.7in} && \nonumber \\
-\frac{2\eta^4}{\g^2}\left[\tilde h_1 f^{n_1} e^{-1/(\tilde b_1 \g^2)}
 - \tilde h_2 f^{n_2} e^{-1/(\tilde b_2 \g^2)} \right] 
\left[\tilde h_1 \tilde b_1^{-1} f^{n_1} e^{-1/(\tilde b_1 \g^2)}
- \tilde h_2 \tilde b_2^{-1} f^{n_2} e^{-1/(\tilde b_2\g^2)}\right]  =0
&& \label{geq}
\eea
\bea
\frac{1}{\rho} (\rho f')' - n^2\frac{(1-v)^2}{\rho^2} f
-\frac{\g^2}{2}(f^2-\g^2)f
\hspace{3.3in} &&\nonumber \\
{}-\eta^2\left[\tilde h_1 f^{n_1} e^{-1/(\tilde b_1 \g^2)}
 - \tilde h_2 f^{n_2} e^{-1/(\tilde b_2 \g^2)} \right] 
\left[\tilde h_1 n_1 f^{n_1-1} e^{-1/(\tilde b_1 \g^2)}
- \tilde h_2 n_2 f^{n_2-1} e^{-1/(\tilde b_2 \g^2)}\right] = 0
&& \label{feq}
\eea
\be
\left(\frac{v'}{\g^2 \rho}\right)'
+ \frac{1-v}{\rho}(f^2 +\eta^2 \g^4) =0
\label{veq}
\ee
where $\tilde h_i = h_i \eta^{-4} \sqrt{\dGS/2}$, 
$\tilde b_i = 2 b_i \eta^2/(3\dGS)$. Hence
$m_F/m_D = (n_1-n_2)\tilde h_i e^{-1/\tilde b_i}$.

\begin{figure*}
\begin{center}
\includegraphics{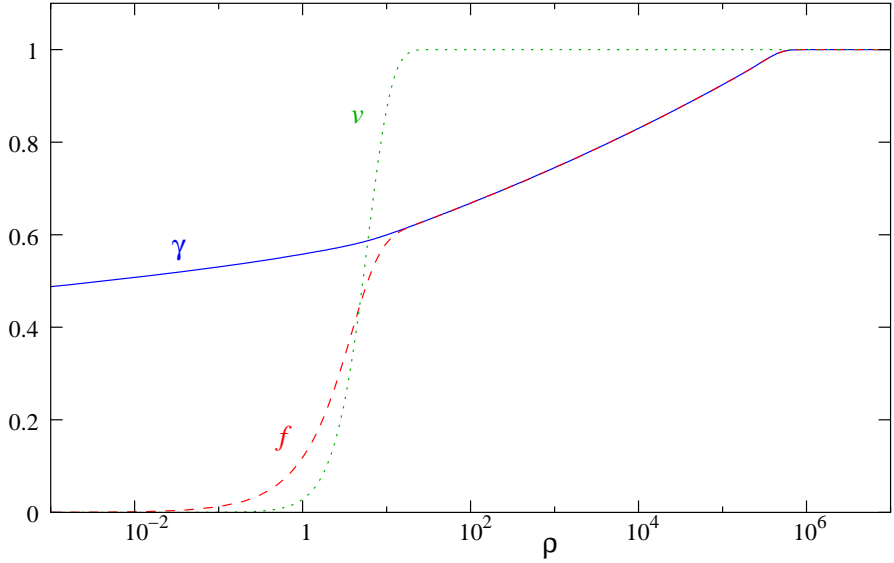}
\caption{\label{fig:gfv}Plot of string profile functions}
\end{center}
\end{figure*}

There are two distinct length scales for this type of string
solution. The string has an inner core of radius $r_D \sim m_D^{-1}$ 
($\rho_D \sim 1$) in which $v < 1$ and $f \neq \g$, 
so $D \neq 0$. Around that region there is an outer core in which 
$v \approx 1$ and $f \approx \g$ but $f,\g < 1$ so $F_0 \neq 0$. This
region is of radius $r_F \sim m_F^{-1}$, which can be far greater then
$r_D$ even for moderate values of the parameters.

A numerical solution of (\ref{geq}--\ref{veq}) for $n=1$ is shown in
figure~\ref{fig:gfv}. The parameters used were $n_1=1$,
$n_2=0$, $g_0=1$, $m_F/m_D = 10^{-5}$ and $b_i = 3N_i/(16\pi^2)$ with
$N_1=4$, $N_2=6$ (as suggested by the symmetry breaking discussed in
section~\ref{sec:model}). The remaining parameters are then $q_0=-2$,
$\dGS=\eta^2=3/(4\pi^2)$, $h_1 \approx 111$ and $h_2 \approx 0.154$.

It is useful to have an approximate analytic solution to the field
equations (\ref{geq}--\ref{veq}). Inside the inner core ($\rho <1$) of
the string the kinetic terms dominate the field equations. At $\rho =1$ we
take $f = \g$ and $v = 1$. Assuming $|n|\eta^2$ is small, the
approximate solution for $\rho <1$ is
\be
\g \approx \frac{A}{\sqrt{1-2|n|\eta^2 A^2\log\rho }}
\label{gsmall}
\ee
\be
f \approx A \rho^{|n|}
\label{fsmall}
\ee
\be
v \approx \frac{\rho^2}{1-2|n|\eta^2 A^2\log\rho} \ .
\label{vsmall}
\ee

If $m_D \gg m_F$ then an approximate solution for the outer core is also
needed. Here $D \approx 0$ so $f \approx \g$. The potential terms in the
field equations can still be neglected, and $v \approx 1$ so some
kinetic terms can also be dropped. Solving eq.~\bref{geq} and imposing
continuity of $\g$ and $\g'$ at $\rho=1$ implies
\be
f \approx \g \approx A \rho^{A^2|n|\eta^2}
\ee
and $v=1$ for $1 < \rho < m_D/m_F$. For $\rho > \rho_F$ we take $f=\g=v=1$.
The constant $A$ is determined by continuity at $\rho = m_D/m_F$. 
It satisfies $(-\log A)/A^2 = |n|\eta^2\log(m_D/m_F)$. For small 
$|n|\eta^2\log(m_D/m_F)$ this implies $A \approx 1-|n|\eta^2\log(m_D/m_F)$.

If $m_D \sim m_F$, there is no outer core. $|\phi|$ and $\sR$ take their
vacuum expectation values for $\rho >1$, and $A=1$.

In all cases the dilaton is thus 
$\sR \approx -2 |n|\dGS \log \rho + \mathrm{const.}$ for small $\rho$. 
Substituting this into the Lagrangian~\bref{LagWZ} we see that
$\mathcal{L} \sim 1/(r \log r)^2$ inside the string core. Thus the
string energy, which is proportional to $\int \mathcal{L} r \, dr$, is
finite for this type of solution. If we had taken $\sR = \mathrm{const.}$ 
then $\mathcal{L} \sim 1/r^2$ and the energy would have diverged
logarithmically at the string core. We see that our solution is a
local string with finite energy, in contrast the usual axion strings
which are global.

The energy per unit length of the string is equal to
\bea
&&\!\!
\mu = 2 \pi \eta^2 \int \rho d\rho \Bigl\{\frac{\g'^2}{\eta^2 \g^2} + f'^2 
+ n^2 \frac{(1-v)^2}{\rho^2} (f^2 + \eta^2 \g^4) 
\nonumber \\ && \hspace{.5in} {}
+ n^2\frac{v'^2}{\g^2 \rho^2}
+\frac{1}{4}(f^2-\g^2)^2 \g^2 
+ \eta^2\left(\tilde h_1 f^{n_1} e^{-1/(\tilde b_1 \g^2)}
-\tilde h_2 f^{n_2} e^{-1/(\tilde b_2 \g^2)}\right)^2 \Bigr\} \ .
\eea
Substituting in the above approximate solution, we find that for small $\eta$, 
\be
\mu = 4\pi n^2 \eta^2\left(2a_n+|n|\eta^2b_n\log\frac{m_D}{m_F}\right)
+O(\eta^4)
\ee
with $a_n,b_n \sim 1$.

\section{Fermion Zero Modes}
\label{sec:fermion}

We will now consider the fermionic sector of the theory. 
The non-zero terms in the fermionic part of the Lagrangian are
\bea
&&
\mathcal{L}_\mathrm{F}=
-i \bar \chi \bar \sigma^\mu 
\left[\partial_\mu - \frac{i}{\sR}(\partial_\mu a -2\dGS A_\mu)\right] \chi
- i \bar \lambda \bar \sigma^\mu 
\left[\partial_\mu + \frac{i}{2\sR}\partial_\mu a \right]\lambda
-i \bar \psi \bar \sigma^\mu \mathcal{D}_\mu \psi
 \nonumber \\ && \hspace{0.4in} {}
-i \bar \psi_0 \bar \sigma^\mu \mathcal{D}_\mu \psi_0
+ \frac{i}{\sqrt{2\sR}}\left(3\frac{\dGS}{\sR} - |\phi|^2\right)
(\bar \lambda \bar \chi - \lambda \chi)
+ \frac{\sR}{2\sqrt{2}}(\bar \chi \bar \sigma^\nu \sigma^\mu \bar
\lambda +\lambda \sigma^\mu \bar \sigma^\nu \chi) F_{\mu \nu}
\nonumber \\ && \hspace{0.4in} {}
- \frac{i\sqrt{2}}{\sqrt{\sR}}
(\bar \phi \lambda \psi - \phi \bar \lambda \bar \psi) 
+ \frac{3}{2}\chi^2 \bar \chi^2
-\left[
\frac{\partial^2 W}{\partial \phi_0 \partial \phi} \psi \psi_0
+2\sR \frac{\partial^2 W}{\partial \phi_0 \partial s} \chi \psi_0 +
(\mathrm{h.\  c.}) \right] \ .
\label{Lagf}
\eea

A common feature of cosmic string models is the existence of fermion
bound states which are confined to the string core. These occur in
models in which fermion fields gain their masses from the cosmic
string Higgs bosons. This is the case for all supersymmetric cosmic
string models, including the one considered in this paper.

Of particular interest are zero energy fermion bound
states, or `zero modes'. Excitations of these are massless particles,
with the unusual property that all excitations of a given zero mode
move in the same direction along the string. The resulting currents
can radically alter the model's cosmology.

\begin{figure*}
\begin{center}
\includegraphics{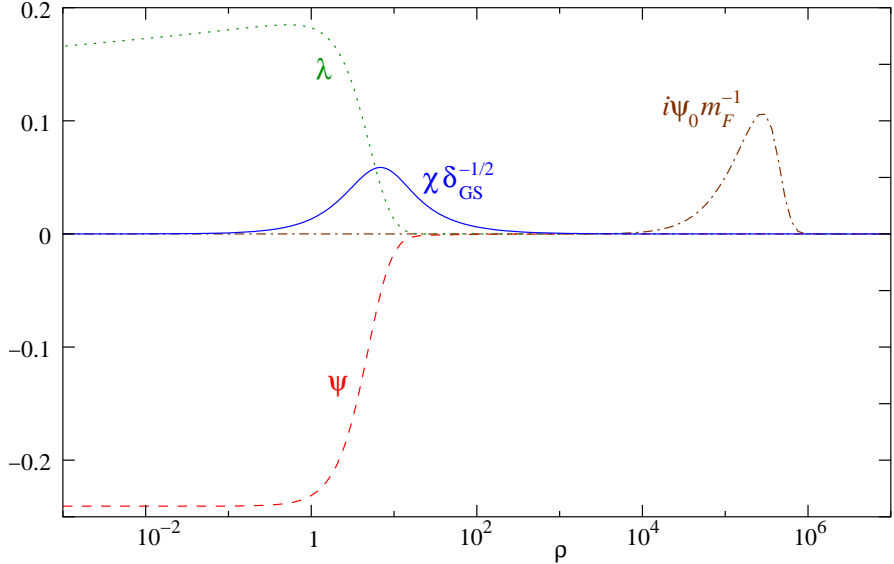}
\caption{\label{fig:zm}Plot of fermion fields for $\epsilon_1 =1$.}
\end{center}
\end{figure*}

In supersymmetric models some of the zero modes can be found by
applying a supersymmetry transformation to the string solution~\cite{sscs}.
Applying the transformation~(\ref{susy1}--\ref{susy3}), with parameter 
$\xi_\alpha = i\sqrt{\dGS}/(2\eta^3)\epsilon_\alpha$, to the string solution
(\ref{ansst}--\ref{ansend}), and using a gauge transformation to
return to Wess-Zumino gauge gives
\bea
\delta \psi_{0 \alpha} &=&  
i\eta \left( \frac{\tilde h_2 f^{n_2}}{e^{1/(\tilde b_2\g^2)}}
- \frac{\tilde h_1 f^{n_1}}{e^{1/(\tilde b_1\g^2)}}
\right) e^{-inq_0\vp} 
\epsilon_\alpha \label{dpsiZ} \ \ \ \ \\
\delta \psi_\alpha &=&  -
\left(f' \pm n\frac{1-v}{\rho}f\right) e^{i(n \mp 1)\vp} \epsilon^\ast_\alpha\\
\delta \lambda_\alpha &=& \left[\frac{\g}{2}\left(\g^2 - f^2\right) 
  	\pm n\frac{v'}{\g \rho}\right] \epsilon_\alpha\label{dlambda} \\
\delta \chi_\alpha &=&  
\left(\frac{\g'}{\g\eta} \mp n\eta \g^2 \frac{1-v}{\rho}\right) e^{\mp i\vp} 
\epsilon^\ast_\alpha\label{dchi}
\eea
where the upper (lower) signs corresponds to the index $\alpha=1$ (2). Since
(\ref{dpsiZ}--\ref{dchi}) are non-zero, supersymmetry is completely
broken inside the string core. The above expressions must be
normalisable if they are to correspond to physical states. At large
$\rho$ they all decay exponentially, so there is no problem there. Using
(\ref{fsmall}--\ref{vsmall}) we see that (\ref{dpsiZ}--\ref{dlambda})
are all well behaved as $\rho \to 0$. The small $\rho$ behaviour of
$\chi_\alpha$ is less obvious. Solving \bref{geq} for small $\rho$ we find
\be
\left(\frac{\g'}{\g \eta} \mp n \eta \g^2 \frac{1}{\rho}\right) = 
\frac{1 \mp \mathrm{sgn} \, n}{-\eta \rho \log \rho} + O(\rho)
 \ee
so $\chi_1$ ($\chi_2$) is well-behaved as $\rho\to 0$ if $n >0$
($n<0$). Thus only one of the transformations ($\xi_1 \neq 0$ if
$n>0$) gives a normalisable fermion zero mode solution. A plot of this
solution is shown in figure~\ref{fig:zm}. Excitations of it are
massless currents, which only flow in one direction along the
string. The direction is determined by the sign of $n$. While the
other transformation also gives a solution of the fermion field
equations, it is unphysical since it is not single valued as $\rho \to 0$.

The total number of zero modes can be determined by
considering solutions to the fermion field equations which are
normalisable at large and small $r$. By trying to match up these
solutions it is possible to find a lower bound for the total number of
physical zero mode solutions. This was done for a general cosmic string
solution in ref.~\cite{Index}. There it was assumed that all fermion
fields had power law behaviour near the centre of the string.
The $\sR^{-1} \partial_\mu a \sim 1/(r \log r)$ terms in
\bref{Lagf} produce extra logarithmic factors in the small $r$
solutions, but these do not affect their normalisability,
so the expression derived in ref.~\cite{Index} still applies. We find
(for $q_0<0$) there are $|n|(1 - q_0)$ zero modes, all of which are
left (right) movers if $n>0$ ($n<0$).

\section{String Network Evolution}
\label{sec:network}

By considering the cosmological properties of the strings, we can
obtain constraints on the underlying theory.
Before doing this, we need to consider the evolution
of a network of our cosmic strings, which may be non-standard. 
For simplicity we will assume that $g_0 \sim 1$. This implies that the
energy per unit length of the string $\mu \sim m_D^2$, and that the
only other mass scales in the model are $\mP$ and $m_F$.

When discussing the evolution of a cosmic string network, it is useful
to introduce the correlation length of the network,
$\xi$. This gives the typical length scale of the network. For
example, in a volume $\xi^3$ there will typically be length $\xi$ of string.

At the phase transition ($T=T_c$) where the strings form, $\xi \sim
1/T_c$. Our strings are essentially formed when the $F$-terms appear in
the potential, at $T=T_c \sim m_F$. This can be far less than the energy
scale of the strings, which is an unusual feature of our model. 

After a brief period of acceleration, the network's evolution is dominated
by the frictional force coming from its interaction with the
plasma. During this friction domination
\be
\xi \sim \frac{\mP^{1/2} m_D}{g_*^{1/4} T^{5/2}} 
\sim g_*^{1/4} t \left(\frac{t}{t_*}\right)^{1/4}
\label{xif}
\ee
(we have used $t \sim \mP/(\sqrt{g_*} T^2)$). This lasts until temperature 
\be
T_* \sim \frac{m_D^2}{\mP} \ ,
\ee  
after which the expansion of the universe (which simply stretches the
strings) will be more significant than the effects of friction. For
our strings, if
\be
\frac{m_F}{m_D} \lesssim \frac{m_D}{\mP} \ ,
\ee
then the friction domination regime will be absent.

For $T < T_*$, the stretching of the strings would soon result in
them dominating the energy density of the universe. This is clearly
not true for our universe, and so some mechanism is needed for the
network to lose energy. In fact, as the strings evolve, loops of
string break off from the network. Typically these decay by radiating
gravitons. As a result of this energy loss the strings do not dominate
the universe and instead the network approaches a scaling solution,
where its energy density scales like matter. The strings move with
relativistic velocities and $\xi =\gamma t$ with $\gamma \sim 1$. The
behaviour of the network in this scaling regime is practically
independent of the network's earlier evolution. We therefore expect
this part of the evolution to be standard for our strings.

If the friction regime is absent or very short, then the string
density at early times will be much higher than that of the scaling
regime. In order for the network to reach a scaling solution, it must
undergo a period of intense loop production. Once the strings have
reached relativistic speeds (which we assume happens rapidly) we find
that during this period the correlation length is given by~\cite{CS}
\be
\xi \approx \gamma t + \xi_c \left(1-\gamma \frac{t_c}{\xi_c}\right) 
\left(\frac{t}{t_c}\right)^{3/4} \ ,
\label{xips}
\ee
which approaches the scaling solution as $t$ increases.

Most of the cosmological constraints on our model arise from loop
production by the string network. 
We will take the typical size of a loop produced by an evolving
network to be $\ell \sim \alpha \xi$. The rate at which loops are produced
is~\cite{CS,loopcoll}
\be
\frac{dn_\mathrm{loop}}{dt} \sim \frac{1}{\xi^3 \ell} 
\sim \frac{1}{\alpha \xi^4}\frac{d\xi}{dt} \ .
\label{dnldt}
\ee
The parameter $\alpha$ depends on the behaviour of the
network. In the friction domination regime  $\alpha \sim 1$. For
strings moving with relativistic velocities (as in the scaling
regime), the value of $\alpha$ is unclear. It is bounded below by
$\alpha \gtrsim \Gamma G\mu$, where the factor
$\Gamma$ is related to how length scales in the network evolve with
time. Numerical simulations suggest $\Gamma \sim 65$.
Simulations of conventional strings indicate $\alpha \sim
10^{-3}$, so we expect large numbers of small loops will be produced
during the scaling regime.

Clearly the scaling regime will not be achieved if the string
loops do not decay. In the previous section we have seen that there
are fermion zero modes in the core of our strings. Including the $z$
and $t$ dependence of the zero modes results in the string carrying a
conserved current~\cite{witten}. Sufficiently large currents will
allow stable string loops, called `vortons', to exist. If these do not
decay the string energy density will come to dominate the universe
after all. This is generically a potential problem for all current
carrying strings. If the zero modes move in both directions the net
current will be smaller than the total number of particles, weakening
the bounds on the model. However in our model we have either left or
right movers, but not both. Thus the net current is maximal. Vorton
bounds can sometimes be avoided by using supersymmetry breaking to
change the fermion spectrum. However, since our zero modes are chiral,
they will survive this~\cite{DDT}, as they are unable to mix with
other modes and become bound states.

Not all loops will go on to become vortons. Only those with a
sufficiently large number of fermion charge carriers will survive. In
the friction dominated regime, the majority of the loops produced when 
$T \sim m_F$ satisfy this requirement. Calculating the typical loop
length and corresponding current allows the vorton density to be
estimated~\cite{CD}
\be
\frac{\rho_v}{g_*T^3} \sim \epsilon \frac{m_F^3}{\sqrt{g_{*F}}\mP m_D}
\label{rhov1}
\ee
where $g_{*F} = g_*(T=m_F)$. For our model, anything in the range  $10
\lesssim g_{*F} \lesssim 10^3$ is possible. The parameter $\epsilon$
measures the efficiency with which a network produces potential
vortons. It will be in the range $0<\epsilon \lesssim 1$.

If $m_F < T_*$ then our strings will form in the
scaling regime, and the above expression~\bref{rhov1} will not apply.
During the scaling regime, far fewer loops will possess enough charge
carriers for them to form vortons. The number density of vortons will
therefore be smaller, as will the typical current per vorton. Taking
these issues into account gives a smaller expression for the vorton
density~\cite{CD}
\be
\frac{\rho_v}{g_* T^3} \sim 
\epsilon \alpha^{3/2} \frac{m_F^2 \mP^2}{\Gamma^2 m_D^3}
\left(\frac{\Gamma m_D}{\sqrt{g_{*F}}\mP}\right)^{\epsilon/(3-\epsilon)} \ .
\label{rhov2}
\ee
The smaller value of $\rho_v$ will weaken any vorton-related constraints
on our model.

The unusual form of the axion strings in this model can reduce the
vorton density even further. Typically the radius of a stable loop
will be of order $10^2 m_D^{-1}$. If $m_F/m_D \lesssim 10^{-2}$ the
outer cores of each side of a potential vorton will overlap. We expect
there to be significant interaction between the charge carriers on the
opposite sides of the loop, allowing the current to decay.  It is
then energetically favourable for the loop to decay, releasing the
parent particles as radiation. This will avoid any constraints on the
model coming from vortons. However other constraints relating to
particle production will apply instead.

\section{Cosmological Constraints}
\label{sec:cosmo}

\subsection{Dark matter bounds}

The existence of conserved currents on cosmic strings has profound
effects on their cosmology. As we discussed in the previous section,
loops of string can be stabilised by the angular momentum of the
charge carriers, producing vortons. If large numbers
of such vortons are produced they can, for example, overclose the
Universe, or interfere with nucleosynthesis. Thus, current-carrying
strings have a vorton problem, similar to the monopole problem of
grand unified theories. In ref.~\cite{BCDT} it was shown that vortons
place very stringent constraints on the underlying particle physics
theory. Indeed, the most stringent constraints arise for chiral
strings (like those in our model) since the current is maximal in this
case~\cite{CD}.

For our model to be consistent with the observed light element
abundances, the universe must be radiation dominated at
nucleosynthesis. This implies $\rho_v \ll g_* T^4$ when 
$T=T_N\sim 10^{-4} \GeV$.

If vortons form during a period of friction domination then $\rho_v$ is
given by eq.~\bref{rhov1} and (taking $\epsilon \sim 1$) 
$\rho_v \ll g_* T_N^4$ implies
\be
g_{*F}^{-1/5} \left(\frac{m_F}{T_*}\right)^{6/5}  G\mu
\ll \left(\frac{T_N}{\mP}\right)^{2/5} \sim 10^{-9} \ .
\label{TNcon1}
\ee

Since the above expression is only valid for $m_F>T_*$, we see that
(assuming the vortons do not decay) the above nucleosynthesis
constraint implies $\dGS \sim G\mu \ll 10^{-9}$. This is a very small
value for $\dGS$, and it is hard to reconcile it with the string
theory motivation for our model. On the other hand, if 
$m_F/m_D \lesssim 10^{-2}$ and the vortons are unstable when their cores
overlap, there is no problem.  For $m_F \gtrsim 10^{-2} m_D$,
eq.~\bref{TNcon1} implies that we must have $\dGS \sim G\mu \ll 10^{-17}$.

If the vortons form during the scaling regime, $\rho_v$ will instead be
given by eq.~\bref{rhov2}. If vortons are unstable for small $m_F$,
then this expression only applies in the limited
parameter range $m_D/\mP \gtrsim m_F/m_D \gtrsim 10^{-2}$. For
$\epsilon=1$ and $\alpha \sim \Gamma G\mu$, this implies $\rho_v/(g_*
T^3) \gtrsim 10^{-9}$. Hence this region of parameter space is ruled
out. If however the vortons do not decay when their
outer cores overlap, the constraint $\rho_v \ll g_* T_N^4$ implies
\be
g_{*F}^{-1/8} (G\mu)^{5/8}  \frac{m_F}{m_D}
\ll \left(\frac{T_N}{\mP}\right)^{1/2} \ .
\label{TNcon2}
\ee
for $\epsilon=1$ and $\alpha \sim \Gamma G\mu$. By taking $m_F$
sufficiently small this can be satisfied for any $m_D$ (even
$m_D=\mP$, in which case $m_F \ll 10^8 \GeV$).

Of course, if $m_F \lesssim 10^{-2} m_D$ and the fermion
currents do decay due to core overlap, the vortons will not be stable.
Then the nucleosynthesis constraint does not apply in the first
place, irrespective of whether the strings form in the scaling regime
or the friction regime.

A similar bound comes from the considering the effects of the dark
matter density on galaxies. An excessive amount of dark matter (this
includes vortons) will lead to problems with galaxy formation. To
avoid this, the vorton density needs to be less than the critical
closure density $\rho_c \approx g_* m_c T^3$, with 
$m_c \sim 1\eV$~\cite{CD}. We find that similar constraints to
eqs.~\bref{TNcon1} and \bref{TNcon2} apply, with $T_N$ replaced by
$m_c$. If the vortons do not decay, we need 
$G\mu \sim \dGS \lesssim 10^{-10}$ if the strings form
during friction domination. If the strings form in the scaling regime,
then $\dGS$ is not bounded if $m_F \lesssim 10^7 \GeV$. Of course, if
the vortons are not stable, the dark matter bound is evaded.

\subsection{Particle production}

If $m_F \lesssim 10^{-2} m_D$ vortons are unlikely to form. Instead the
loops will decay, although in contrast to more conventional models, the
main decay mechanism will be particle production rather than
gravitational radiation. 

The dilaton couples to our axions strings with a similar strength
to gravity, and so decaying string loops will radiate not only
gravitons but also dilatons~\cite{dilacons}. Furthermore, when the
radius of a loop is comparable to its thickness, microphysical forces
will become important, and all the energy of the loop will be released
as a burst of radiation~\cite{loopcoll}. 

We will start by finding an expression for the number of dilatons
emitted by a loop as it shrinks. For simplicity we
will only consider strings in the scaling regime, where frictional
forces can be neglected, and $\ell \sim \alpha t$. We follow the
method of  ref.~\cite{dilacons}. There it was assumed that 
$\ell \ll \ell_\mathrm{crit} = 4\pi/m_F$ ($m_F$ is the dilaton mass
for our model). This will not be the case for our strings, although
the analysis is easily extended to general $\ell$. The dilaton number
density is then
\be
\frac{n_F}{s} \sim \left( \frac{\alpha \mP}{\sqrt{g_{*F}} m_F }\right)^{1/2}
G\mu \times \left\{ \begin{array}{ll}
 (\ell m_F)^{1/2} & \ell \ll \ell_\mathrm{crit}\\
 (\ell m_F)^{-5/6} & \ell \gg \ell_\mathrm{crit} \end{array} \right. \ .
\label{dc1}
\ee
where $s \sim g_* T^3$ is the entropy density. We see that the maximum
contribution comes from loops with $\ell \sim \ell_\mathrm{crit} \sim m_F^{-1}$.

Since the string network does not exist before $t_c$, the
minimum value of $\ell$ is $\alpha t_c$, which can be greater than
$\ell_\mathrm{crit}$. Hence
\be
\frac{n_F}{s} \sim
\left(\frac{\sqrt{g_{*F}}m_F}{\alpha \mP}\right)^{1/3} G\mu
\label{dc2}
\ee
when $m_F < \alpha \mP/\sqrt{g_{*F}}$. This leads to weaker
constraints than the $\ell = \ell_\mathrm{crit}$ results of
ref.~\cite{dilacons}.

The derivation of the above expression assumed that the behaviour of
the network was given by a scaling solution. This is not the case for our
strings, since if $m_D \sim m_F$ there will also be the friction
regime. Even if $m_F \ll m_D$ and the friction regime is absent, the
initial behaviour of the network (which is when most dilatons are
produced) will instead be given by eq.~\bref{xips}. The string density
was higher at this time, and so we expect more dilatons to be
produced. The actual bound for our model will be somewhere between the
two expressions \bref{dc1} and \bref{dc2}.

If the dilatons have a lifetime of less than 0.1s, they will decay
well before nucleosynthesis, and there will be no significant bounds
on the model from dilaton production. This will be the case if 
$m_F \gtrsim 10^2 \TeV$. For a lighter dilaton, with 
$m_F \sim 10^4 - 10^2 \GeV$, we need $n_F/s \lesssim
10^{-14}$~\cite{dilacons}.  Taking $\alpha \sim \Gamma G \mu$, the
bound~\bref{dc2} implies  $G\mu \lesssim 10^{-11}
(\TeV/m_F)^{1/2}$. Alternatively if we use the bound~\bref{dc1} with
$\ell = \ell_\mathrm{crit}$, we have instead 
$G\mu \lesssim 10^{-15} (m_F/\TeV)^{1/3}$. Either way a very small
value of $m_D$ is required to satisfy the constraints for light dilatons.

Since the dilaton is one of the particles which our strings are made up
of, our model has an additional dilaton production mechanism. When the
loop size is comparable to its width, it will collapse and release its
energy as a burst of radiation. This will be consist of dilatons and
all the other particles that the strings are made of.

Using eq.~\bref{dnldt} and following Brandenberger {\em et
al.}~\cite{loopcoll}, the total number of quanta produced in loop
collapse from a string network is
\be
n_Q(t) \sim \frac{N_Q}{a(t)^3} \int^\xi_{\xi_c} 
\frac{a(t')^3}{\alpha \, \xi(t')^4} d\xi'
\ee
where $N_Q$ is the number of quanta produced by a single collapsing loop.

At the time of decay, the length of loop will be about one order of magnitude
greater than the width of the outer core, i.e.\ $\ell \sim \beta m_F^{-1}$,
with $\beta \geq 2\pi$. We expect to get $N_F = \ell m_F \sim \beta \sim 10$
particles of mass $m_F$ from the outer core and $N_D = \ell m_D \sim \beta
m_D/m_F$ from the inner core.

In the friction dominated regime $\xi \sim t^{5/4}$,
and so~\cite{loopcoll} 
\be
\frac{n_Q}{s} \sim \frac{N_Q}{\alpha g_{*F}}
\label{nq2}
\ee
(with $\alpha \sim 1$).
If $g_{*F} \sim 10^3$, as would be the case if $m_F$ is around the GUT
scale, $n_F/s \sim N_F/g_{*F} \sim 10^{-2}$ and $n_D \sim (m_D/m_F) n_F$.

An equivalent calculation in the scaling regime would give smaller
values of $n_Q$. However the early evolution of our strings is not
actually a scaling solution, and it is the early evolution which
produces the largest number of loops. Using eq.~\bref{xips} we find
that the above expression~\bref{nq2} still applies, although this time
$\alpha$ is much smaller, and so the number of particles produced is
very high.  Unless the lifetime of the dilatons is less than
0.1s, this is completely incompatible with nucleosynthesis
constraints~\cite{dilacons}. This imposes a lower bound on $m_F$ of $10^2\TeV$.

\subsection{Microwave background constraints}

Finally, if the theory evades the vorton bounds then the string
network will give rise to anisotropies in the cosmic microwave
background. Experimentally these are detected at the level of about
$10^{-6}$. In the cosmic string scenario for structure formation they
are of order $G\mu$. So unless $g_0^2 \dGS \lesssim 10^{-7}$ they
are too large and can be ruled out by the COBE result. This bound
comes from considering the properties of the late time scaling
solution of the network. Since this is independent of the non-standard
early behaviour of our strings, it is hard to see how this bound can
be avoided. It is hard to reconcile such a small value of $\dGS$ with
the underlying theory which motivates our model. If $s$ were a moduli
field rather than the dilaton, a smaller value of $\dGS$ could be possible.

It should be noted that, throughout our analysis, we have assumed that
our strings intercommute (i.e.\ when two strings collide, the
different halves of the two strings exchange partners). For the usual
$U(1)$ strings this is only true if the gauge mass is smaller than the
Higgs mass. It is not clear if our strings satisfy an equivalent
relation. If they do not intercommute, then a network of our strings
will not produce loops, and will evolve in a significantly different
way. The microwave background bound will also be changed.
However if loops do not break off from the string network, then
it is hard to see how it can reach a scaling solution and
avoid dominating the universe. Thus this kind of non-standard
evolution could actually produce even stronger constraints on the model.

\section{Bogomolnyi equations}
\label{sec:bog}

It is actually possible to reduce the field
equations of the model~(\ref{seq}--\ref{Aeq}) to first order
differential equations by using the symmetries of the model. 
Using the $z$ and $t$ independence of the fields, $\phi_0=0$, and
integration by parts, the integral of the Lagrangian \bref{LagWZ} can
be rewritten as
\bea
\mu_1 =&& \nsp \int d^2x \Biggl\{
|(\partial_x - i A_x)\phi \pm i(\partial_y - i A_y)\phi|^2 
\nonumber \\ && {} 
+{1 \over 4 \sR^2}
|(\partial_x s - 2i\dGS A_x) \pm i(\partial_y s - 2 i \dGS A_y)|^2
+ {\sR \over 2} \left(F_{xy} \pm D\right)^2  + |F_0|^2 \Biggr\}\ .
\eea
We use the upper (lower) sign for $n > 0$ ($n <0$). If $F_0=0$ and
\be
f' = |n| \frac{1-v}{r} f
\label{bphieq}
\ee
\be
\sR' = 2\dGS |n| \frac{1-v}{r}
\label{bseq}
\ee
\be
|n|{v' \over r} = \frac{1}{\sR}\left({\dGS \over \sR} - f^2\right)
\label{bAeq}
\ee
then $\mu_1=0$.
Since $F_0=0$ implies $s = 2\dGS\ln (\phi/\eta) + \dGS/\eta^2$,
eq.~\bref{bphieq} is consistent with eq.~\bref{bseq}. Although the
above equations give a solution of the full field equations, the
string solution has $\sR$ passing through zero. This implies the
coupling $g \to \infty$ there, and so the solution is unphysical.

\section{Conclusions}
\label{sec:con}

In this paper we have investigated a non-standard class of axionic
cosmic string solutions. The axion model used was motivated by an
effective superstring action. In this type of model the axion
and dilaton form part of the same complex field. By allowing the
dilaton to vary, and by including a gauge field in the model, it was
possible to find finite energy cosmic string solutions without having
to introduce cutoffs. This contrasts with the usual axion string
models, in which the dilaton is absent or constant. 

Another unusual
feature of our strings is the existence of two length scales for the
size of the string core. The string's magnetic field is confined to
an inner core, while the dilaton (and a Higgs field) can vary over a
far greater distance. The two length scales arise from the two mass
scales in the theory's potential. One ($m_D$) comes from high energy
supergravity contributions, and the other ($m_F$) from a lower energy gaugino
condensate, which is used to stabilise the dilaton.

Being supersymmetric, our model also contains fermion fields. As is
common with topological defects in supersymmetric models, our strings
possess zero energy fermion bound states. Excitations of
these form conserved chiral currents on the strings. These currents
can allow stable string loops (vortons) to exist. The existence of
vortons suggests extremely strong cosmological constraints on the
model.

However, the fact that the strings have two mass scales alters their
evolution. By taking $m_F$ to be small, the strings form at a later time
with a smaller density. This leads to far weaker vorton constraints
than those that would be expected for the string's large energy per
unit length ($m_D^2$). If $m_F/m_D$ is sufficiently small the outer
cores of the opposite sides of the vorton will overlap. Since the
fermion bound state wavefunctions do reach to the outer core, this
suggests the currents will decay, destabilising the vortons and
evading the corresponding constraints on the strings.

If string loops are unstable there will be other constraints on the
model from particle production by the strings. They
will radiate dilatons as they shrink (in addition to
gravitational radiation). Furthermore, when the loops collapse, they
will decay into yet more dilatons (as well as heavy Higgs and gauge
particles). The number of dilatons produced by these
processes is very high, and will rule out the model unless the dilaton
lifetime is short. This imposes a lower bound of $10^2 \TeV$ on the
dilaton mass, which is $m_F$ for our model. This rules out the gluino
condensation as the source of the potential which stabilises the dilaton. 
Combining the above bounds gives $10^{-2} m_D \gtrsim m_F \gtrsim 10^2 \TeV$.

If $m_F \ll m_D$ the strongest constraints on the model will come from
cosmic microwave background measurements. These suggest 
$m_D^2/\mP^2 \sim \dGS \lesssim 10^{-7}$. Although weaker than the
vorton bounds, this is still difficult to reconcile with the string
theory motivation for the model. This would be a less serious problem
if we took the superfield $S$ to be a moduli field rather than the
dilaton. Thus this type of cosmic string could still be both
theoretically and observationally viable. To satisfy all the above
bounds, we require $10^{15}\GeV \gtrsim m_D \gtrsim 10^7\GeV$.

An important assumption in the microwave background constraints is
that the strings intercommute. This may not be the case for our
strings, in which case a string network will not produce loops or
approach a scaling solution in the usual way. This would significantly
alter the bounds on the model, although unfortunately we expect it to
make them much stronger.

\section{acknowledgements}
SCD thanks C. J. A. Martins and E. P. S. Shellard for helpful
comments. We are all grateful for
a Royal Society-CNRS exchange grant which enabled us to pursue this 
investigation. SCD is grateful to PPARC, the EU network
HPRN--CT--2000--00152, and the Swiss Science Foundation for financial
support. ACD also thanks PPARC.

\appendix
\section{Conventions}

We use the metric $\eta_{\mu \nu} = \mbox{diag}(1,-1,-1,-1)$.

Under the U(1) gauge symmetry the fields transform as follows
\bea
&& \Phi \rightarrow e^{2i\Lambda}\Phi \ , 
\ \ S \rightarrow S + 4i\dGS \Lambda \ ,
\nonumber \\ &&{} 
V \rightarrow V + i (\Lambda - \Lambda^\dagger) \ , 
\ \ \Phi_0 \rightarrow e^{-2i q_0 \Lambda}\Phi_0
\eea
so $q_\phi = -1$.

Under the R symmetry
\be
S \rightarrow S + 2b_0i\alpha \ , \ \ \Phi \rightarrow e^{-i r_\phi \alpha}\Phi
\ , \ \ \Phi_0 \rightarrow e^{-i r_0 \alpha} \Phi_0
\ee

Under a supersymmetry transformation of the bosonic fields, the change
in the fermion fields (in Wess-Zumino gauge) is
\be
\delta \psi = \sqrt{2} F \xi
	+ i\sqrt{2} \sigma^\mu \bar \xi \mathcal{D}_\mu \phi
\label{susy1}
\ee
\be 
\sR^{-1/2} \delta \lambda = iD \xi
	+ {1 \over 2} \sigma^\mu \bar \sigma^\nu \xi F_{\mu \nu}
\ee
\be
2\sR \delta \chi = \sqrt{2} F_s \xi 
	+ i\sqrt{2} \sigma^\mu \bar \xi (\partial_\mu s - 2i \dGS A_\mu)
\label{susy3}
\ee

\newcommand{\bt}[1]{}


\begin{thebibliography}{99}
\bibitem{supstr}
M.~Dine, N.~Seiberg and E.~Witten,
\bt{Fayet-Iliopoulos Terms In String Theory,}
Nucl.\ Phys.\ B {\bf 289} (1987) 589.
J.~J.~Atick, L.~J.~Dixon and A.~Sen,
\bt{String Calculation Of Fayet-Iliopoulos D Terms In Arbitrary Supersymmetric
Compactifications,}
Nucl.\ Phys.\ B {\bf 292} (1987) 109.
M.~Dine, I.~Ichinose and N.~Seiberg,
\bt{F Terms And D Terms In String Theory,}
Nucl.\ Phys.\ B {\bf 293} (1987) 253.
\bibitem{GS} 
M.~B.~Green and J.~H.~Schwarz,
\bt{Anomaly Cancellation In Supersymmetric D=10 Gauge Theory And Superstring
Theory,}
Phys.\ Lett.\ B {\bf 149} (1984) 117.
\bibitem{CS}
C.~J.~A.~Martins and E.~P.~S.~Shellard,
\bt{Extending the velocity-dependent one-scale string evolution model,}
Phys.\ Rev.\ D {\bf 65}, 043514 (2002) [hep-ph/0003298]. 
A.~Vilenkin and E.~P.~S.~Shellard, {\em Cosmic Strings
and Other Topological Defects}, Cambridge University Press (1994). 
\bibitem{oldaxion} 
J.~A.~Harvey and S.~G.~Naculich,
\bt{Cosmic Strings From Pseudoanomalous U(1)S,}
Phys.\ Lett.\ B {\bf 217} (1989) 231. 
J.~A.~Casas, J.~M.~Moreno, C.~Munoz and M.~Quiros,
\bt{Cosmological Implications Of An Anomalous U(1): Inflation, Cosmic Strings
And Constraints On Superstring Parameters,}
Nucl.\ Phys.\ B {\bf 328} (1989) 272.
\bibitem{ax1}
P.~Binetruy, C.~Deffayet and P.~Peter,
\bt{Global vs. local cosmic strings from pseudo-anomalous U(1),}
Phys.\ Lett.\ B {\bf 441} (1998) 52 [hep-ph/9807233].
\bibitem{Dstrings}
G.~Dvali, R.~Kallosh and A.~Van Proeyen,
\bt{D-term strings,}
JHEP {\bf 0401}, 035 (2004)
[hep-th/0312005].\\
G.~Dvali and A.~Vilenkin,
\bt{Formation and evolution of cosmic D-strings,}
JCAP {\bf 0403}, 010 (2004) [hep-th/0312007]. 
E.~J.~Copeland, R.~C.~Myers and J.~Polchinski,
\bt{Cosmic F- and D-strings,}
JHEP {\bf 0406}, 013 (2004) [hep-th/0312067].
\bibitem{krasnikov}
N.~V.~Krasnikov,
\bt{On Supersymmetry Breaking In Superstring Theories,}
Phys.\ Lett.\ B {\bf 193} (1987) 37.
\bibitem{sscs}
S.~C.~Davis, A.~C.~Davis and M.~Trodden,
\bt{N = 1 supersymmetric cosmic strings,}
Phys.\ Lett.\ B {\bf 405} (1997) 257 [hep-ph/9702360].
\bibitem{witten}
E.~Witten, 
\bt{Superconducting strings,} Nucl.\ Phys.\ {\bf B249} (1985) 557.
\bibitem{Index} 
S.~C.~Davis, A.~C.~Davis and W.~B.~Perkins,
\bt{Cosmic string zero modes and multiple phase transitions,}
Phys.\ Lett.\ B {\bf 408} (1997) 81 [hep-ph/9705464].
\bibitem{DDT}
S.~C.~Davis, A.~C.~Davis and M.~Trodden,
\bt{Cosmic strings, zero modes and SUSY breaking in nonabelian N = 1 gauge
theories,}
Phys.\ Rev.\ D {\bf 57} (1998) 5184 [hep-ph/9711313].
\bibitem{CD}
B.~Carter and A.~C.~Davis,
\bt{Chiral vortons and cosmological constraints on particle physics,}
Phys.\ Rev.\ D {\bf 61} (2000) 123501 [hep-ph/9910560].
\bibitem{BCDT}
R.~H.~Brandenberger, B.~Carter, A.~C.~Davis and M.~Trodden,
\bt{Cosmic vortons and particle physics constraints,}
Phys.\ Rev.\ D {\bf 54} (1996) 6059 [hep-ph/9605382].
\bibitem{dilacons}
T.~Damour and A.~Vilenkin,
\bt{Cosmic strings and the string dilaton,}
Phys.\ Rev.\ Lett.\  {\bf 78} (1997) 2288 [gr-qc/9610005]. 
T.~Damour, F.~Piazza and G.~Veneziano,
\bt{Violations of the equivalence principle in a dilaton-runaway scenario,}
[hep-th/0205111].
\bibitem{loopcoll}
R.~H.~Brandenberger, A.~C.~Davis and M.~Hindmarsh,
\bt{Baryogenesis from collapsing topological defects,}
Phys.\ Lett.\ B {\bf 263} (1991) 239.
\end{thebibliography}
\end{document}